\documentclass[conference,letterpaper]{IEEEtran}
\IEEEoverridecommandlockouts
\usepackage{amsmath,amssymb}
\usepackage{graphicx,booktabs,array,cite,balance}
\usepackage{xcolor,tikz}
\usetikzlibrary{arrows.meta,positioning,calc}
\usepackage[hidelinks]{hyperref}

\newcommand{\Pb}{P_\mathrm{Bat}}
\newcommand{\Cb}{C_\mathrm{Bat}}
\newcommand{\tc}{\mathrm{Cur}}

\usepackage{xurl}

\newcommand{\resultplot}[3]{%
  \IfFileExists{#1}{\includegraphics[width=\columnwidth]{#1}}{%
  \fbox{\begin{minipage}[c][#2][c]{0.94\columnwidth}\centering
  \color{blue}\small\textbf{PSCAD results to be inserted}\\[5pt]#3
  \end{minipage}}}}

\title{Low-Voltage Ride-Through in AI Data Centers via Coordinated Battery and Computing Load Control}

\author{\IEEEauthorblockN{Xiaoyang Wang, Xin Chen, Fang Chen Lin, Prasad Enjeti}
\IEEEauthorblockA{\textit{Department of Electrical and Computer Engineering} \\
\textit{Texas A\&M University}, College Station, TX, USA \\
\{wangxy, xin\_chen, fclin3, enjeti\}@tamu.edu}
\and
\IEEEauthorblockN{Ali Yazdanpanah, Yunzhi Cheng, Shri Abhyankar}
\IEEEauthorblockA{\textit{Electric Reliability Council of Texas (ERCOT)} \\
Taylor, TX, USA
\\
\{Ali.yazdanpanah,Yunzhi.Cheng, shri.abhyankar\}@ercot.com}
\thanks{This work was supported in part by NSF CAREER 2541998, in part by  the Electric Reliability Council of Texas (ERCOT) through its Grid Research, Innovation, and Transformation (GRIT) initiative under Research Agreement Phase II, and in part by the Power Systems Engineering Research Center (PSERC) Project S-118. }
}

\begin{document}
\maketitle

\begin{abstract}
The ability of large AI data centers to ride through grid voltage sags is critical to power system stability and reliability. 
This paper frames data center low-voltage ride-through (LVRT) as a DC-link power balance problem. Once the rectifier reaches its current limit, the power drawn from the grid decreases in proportion to the grid voltage. The DC-link capacitor can supply the resulting power deficit only briefly before the DC-link voltage falls low enough to trigger undervoltage protection tripping.
To address this issue, this paper proposes effective control strategies for coordinating uninterruptible power supply (UPS) battery and computing load curtailment to offset the power deficit and enable LVRT at minimum control cost. Three control scenarios are studied: UPS battery support alone, computing load curtailment alone, and the coordinated use of both resources. Electromagnetic transient (EMT)
simulations using the TAMU-ERCOT AI data center model demonstrate that battery support and the coordinated strategy
achieve LVRT for all tested sags, including a near-zero-voltage
fault, while load curtailment alone rides through moderate voltage sags.
\end{abstract}


\section{Introduction}
As large data center loads continue to grow, their ability to ride through voltage disturbances has become important for bulk power system reliability \cite{CHEN2026100162}. Following a normally cleared transmission fault in July 2024, the North American Electric Reliability Corporation (NERC) documented approximately 1,500 MW of simultaneous data center load reduction \cite{nercIncident}. The Electric Reliability Council of Texas (ERCOT) also identified 26 events since 2023 in which data center or crypto-mining loads tripped offline during normal voltage disturbances \cite{ercotBoard282}. 
Such sudden losses of large loads
can cause frequency and voltage
excursions that threaten grid stability~\cite{nercIncident}.

In response, ERCOT's Nodal Operating Guide Revision Request (NOGRR) 282
specifies voltage ride-through (VRT) requirements for large
computational loads~\cite{ercot282}.
Denmark’s transmission system operator, Energinet, has issued VRT requirements for transmission-connected demand facilities~\cite{energinetTR343}, and Finland’s transmission system operator, Fingrid, has proposed similar requirements~\cite{fingridKJV2026}. 
Table~\ref{tab:vrt} presents ERCOT's requirements on the minimum duration for which a large load must
remain connected for each voltage band at the point of common coupling
(PCC). In addition to remaining connected, the load must keep drawing its
pre-disturbance power when the PCC voltage is between 0.9 and 1.1~p.u., may reduce its active power consumption in proportion to the PCC voltage between 0.5 and 0.9~p.u., and may cease drawing power below 0.5~p.u. After
the PCC voltage recovers above 0.9~p.u., the load must restore to
90\% of its pre-disturbance power within 2~s~\cite{ercot282}.

\begin{table}[t]
\caption{ERCOT NOGRR~282 low-voltage ride-through durations~\cite{ercot282}}
\label{tab:vrt}
\centering\small
\begin{tabular}{cc}
\toprule
RMS voltage at PCC (p.u.) & Minimum duration (s)\\
\midrule
$0.90\leq V_\mathrm{Grid}\leq1.10$ & Continuous\\
$0.80\leq V_\mathrm{Grid}<0.90$ & 2.0\\
$0.50\leq V_\mathrm{Grid}<0.80$ & 0.5\\
$0.35\leq V_\mathrm{Grid}<0.50$ & 0.25\\
$V_\mathrm{Grid}<0.35$ & 0.15\\
\bottomrule
\end{tabular}
\end{table}

Existing studies~\cite{xieVRT,wangSolVRT,norouziVoltageRideThroughLarge2026} primarily frame data center VRT as a voltage regulation problem, using flexible on-site energy resources to maintain voltages in the facility’s distribution network near their nominal values during a sag.  
Formal synthesis has been used to design controllers that regulate active and reactive power to keep the internal bus voltage above its trip threshold while satisfying the power-floor and recovery requirements~\cite{wangSolVRT}. Reactive-power compensation has also been proposed to raise AC voltage during a voltage sag~\cite{norouziVoltageRideThroughLarge2026}.
However, data center VRT differs from conventional voltage regulation. Its objective is to meet grid-code requirements for continued connection and active power draw over prescribed durations under specified voltage disturbances at the PCC, while respecting internal equipment and protection limits.
In particular, the power converter controls and protection logic in the uninterruptible power supply (UPS) system are critical to the facility’s VRT response \cite{eatonGoodGridCitizen}.

\textbf{Contributions}. First,
this paper frames data center low-voltage ride-through (LVRT) as a power balance problem at the UPS DC link. During a voltage sag, the reduced AC input voltage and the rectifier’s current limit constrain the grid power delivered to the DC link, while Information Technology (IT) computing load typically remains near its pre-disturbance level. The resulting power deficit discharges the DC-link capacitor, lowering its voltage and triggering undervoltage protection if the voltage falls below the trip threshold. Second, to address this imbalance, we propose effective LVRT control strategies that coordinate the UPS battery and IT load curtailment to compensate for the power deficit and maintain the DC-link voltage above the trip threshold. The proposed control minimizes the total control cost and is suitable for practical implementation in data centers. Third, high-fidelity electromagnetic transient (EMT) simulations of an AI data center in PSCAD demonstrate that battery support and the coordinated strategy achieve LVRT for all tested sags, including a near-zero-voltage fault, while load curtailment alone rides through moderate sags.

\section{Problem Formulation of LVRT}

\subsection{System Description}

\begin{figure}[t]
\centering
\includegraphics[width=0.95\linewidth]{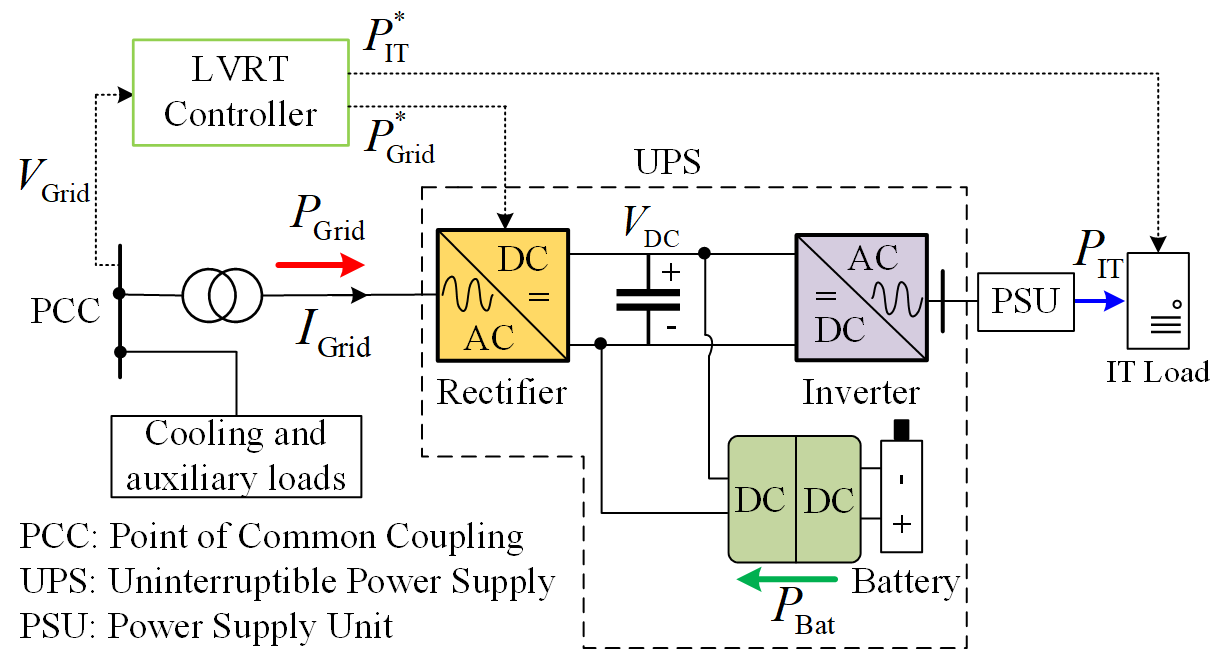}
\caption{Illustration of the power flow paths in a double-conversion UPS-based AI data center and the proposed LVRT control strategy. 
}
\label{fig:system}
\end{figure}

As illustrated in Fig.~\ref{fig:system}, the IT load of an AI data
center is supplied through a typical double-conversion UPS. The
rectifier, connected to the PCC through a transformer, feeds the DC
link, from which the inverter supplies the servers through their power
supply units (PSUs), and the UPS battery is connected to the DC link
through a bidirectional DC/DC converter. Cooling and other auxiliary
loads are connected at the PCC, with or without their own UPS. This
paper focuses on the LVRT of the IT load, which is supplied through the
UPS and can be curtailed readily.

The energy stored in the DC-link capacitor is $E = \frac{1}{2} C V_{\mathrm{DC}}^2$, where $C$ is the capacitance and $V_\mathrm{DC}$ is the DC-link voltage. Neglecting power losses, the DC-link power balance yields the voltage dynamics \eqref{eq:Cdynamic}: 
\begin{equation}
\label{eq:Cdynamic}
 \dot{E}= C V_\mathrm{DC} \dot{V}_\mathrm{DC}=P_\mathrm{Grid} + P_\mathrm{Bat}-P_\mathrm{IT},
\end{equation}
where $P_\mathrm{Grid}$ and $P_\mathrm{IT}$ denote the power drawn from the grid and the IT computing load power, respectively. $P_\mathrm{Bat}$ denotes the UPS battery power output. The grid-side power $P_\mathrm{Grid}$ is:
\begin{equation} \label{eq:gridP}
    P_\mathrm{Grid}=V_\mathrm{Grid} I_\mathrm{Grid}.
\end{equation}
Here, $V_\mathrm{Grid}$ and $I_\mathrm{Grid}$ denote the RMS grid-side voltage and current at the PCC. The voltage and current are assumed to be in phase under the rectifier’s unity power factor control. 
During normal operation with $P_\mathrm{Bat}=0$, the rectifier regulates the DC-link voltage by adjusting the grid-side input current $I_\mathrm{Grid}$ so that $P_\mathrm{Grid}=P_\mathrm{IT}$ at steady state.

\subsection{Voltage Sag-Induced Power Deficit and LVRT}

During a grid voltage sag, the boost capability and voltage regulation of the multistage power conversion chain can maintain a nearly constant supply voltage for the IT load. For a fixed computing workload, the IT load continues to draw approximately constant power $P_\mathrm{IT}$. To regulate the DC-link voltage as the PCC voltage $V_\mathrm{Grid}$ falls, the rectifier increases the grid-side current $I_\mathrm{Grid}$ up to its limit $I_\mathrm{Grid}^\mathrm{max}$ (e.g., 1.2 p.u.). Once the current limit is reached, the maximum active power the rectifier can draw from the grid decreases in proportion to the PCC voltage:
\begin{equation}
    P_\mathrm{Grid}^{\max}(t)=I_\mathrm{Grid}^\mathrm{max}V_\mathrm{Grid}(t).
\end{equation}
Under a severe voltage sag with $P_\mathrm{Grid}^{\max}<P_\mathrm{IT}$, the DC-link capacitor discharges to supply the power deficit, causing $V_\mathrm{DC}$ to decline according to \eqref{eq:Cdynamic}. As the voltage falls, the nearly constant-power IT load draws more current from the DC link. If $V_\mathrm{DC}$ falls below the trip threshold, undervoltage protection can cause the UPS to cease drawing power from the grid.

For a fixed uncompensated power deficit $\Delta P$:
\begin{equation}
\label{eq:powerdeflict}
\Delta P(V_\mathrm{Grid}) = \max\{ P_\mathrm{IT,0}-I_\mathrm{Grid}^{\max}V_\mathrm{Grid}; 0 \},
\end{equation}
using \eqref{eq:Cdynamic},
the hold time for $V_\mathrm{DC}$ to fall from its initial value $V_\mathrm{DC,0}$ to the trip threshold $V_\mathrm{DC,\min}$ can be estimated as:
\begin{equation}
 T_{\rm hold}=  \ \frac{C(V_{\rm DC,0}^2-V_{\rm DC, \min}^2)}{2}  \frac{1}{\Delta P}.
 \label{eq:holdup}
\end{equation}
To enable LVRT, additional power support is needed to prevent an undervoltage trip during the hold time interval $T_\mathrm{hold}$\footnote{For time-delayed undervoltage protection, a brief drop below $V_\mathrm{DC,min}$ may be tolerated if the DC-link voltage recovers sufficiently to reset the timer before the trip delay expires.}.

During a significant voltage sag (e.g., $V_{\mathrm{Grid}} <0.8$ p.u.), conventional UPS protection can prevent DC-link voltage collapse by directly tripping the rectifier, leading to $P_\mathrm{Grid}=0$ and transferring the entire IT load to battery backup. This transfer causes a sudden reduction in grid load approximately equal to the IT power demand $P_\mathrm{IT}$. For a large data center, this load reduction can reach hundreds of megawatts and threaten grid voltage and frequency stability.

Hence, LVRT requires local support to offset the power deficit $\Delta P$ during a grid voltage sag. After the grid voltage recovers, the facility withdraws this support as the grid power draw returns to its pre-disturbance level. Two power resources readily available in an AI data center can be leveraged to offset the power deficit for LVRT: (1) the UPS battery, which can rapidly deliver high discharge power $P_{\mathrm{Bat}}$ and can typically sustain the full IT load for 1-5~min \cite{greenGridUPS}, well beyond the ride-through requirements in Table~\ref{tab:vrt}; and (2) the IT computing load $P_{\mathrm{IT}}$, which can be curtailed 
during the brief ride-through interval, e.g., by GPU power capping or temporary job pausing, with limited impact on service quality~\cite{colangeloEmerald}.

\section{LVRT Control Design for UPS
Battery and Computing Load Coordination}
\label{sec:Scenario}
In this section, we develop effective LVRT control strategies for practical implementation in AI data centers.  
 We consider three scenarios: UPS battery support alone, computing load curtailment alone, and the coordinated use of both resources.

\subsection{Grid-Power Reference and Power Deficit Calculation}

\begin{figure}[htbp]
    \centering
    \vspace{-7pt}
\includegraphics[width=0.8\linewidth]{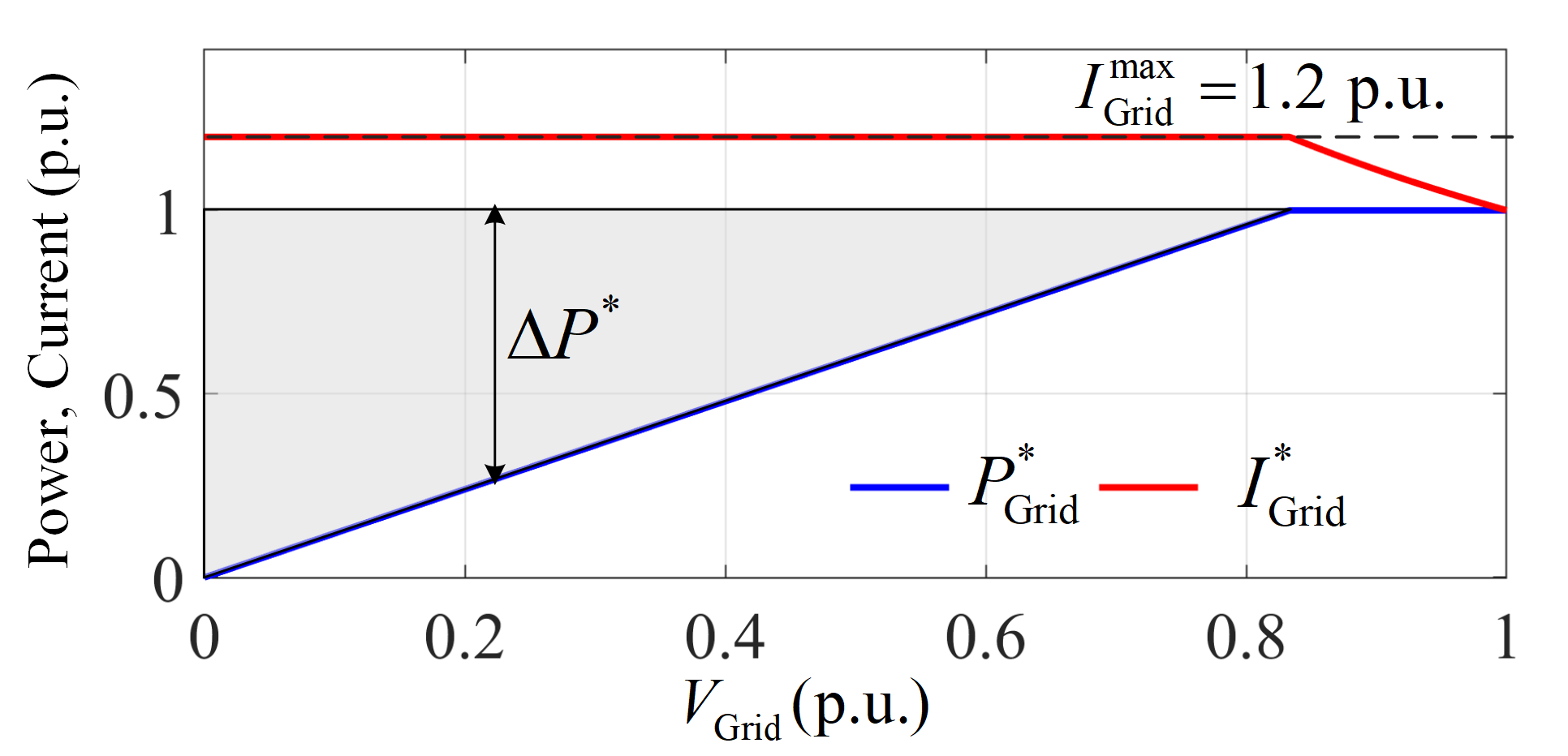}
   \vspace{-5pt}
    \caption{Grid power reference $P_\mathrm{Grid}^*$ and rectifier current $I_\mathrm{Grid}$ as functions of grid voltage $V_\mathrm{Grid}$, with $P_\mathrm{IT,0}=1$~p.u. and $I_\mathrm{Grid}^{\max}=1.2$~p.u., along with the power deficit $\Delta P^*$.
    }
    \label{fig:GridPowerControl}
\end{figure}

In Scenarios~1 and~3, the rectifier operates in constant-power control during a sag and tracks a voltage-dependent active power reference $P_\mathrm{Grid}^*$. Given the grid voltage $V_\mathrm{Grid}(t)$, we set this reference to the maximum power the rectifier can draw without exceeding its current limit, given by
\begin{align}
\label{eq:Pgridobj}
P_\mathrm{Grid}^*\big(V_\mathrm{Grid}(t)\big) &=\min \big\{P_\mathrm{IT,0},  I_\mathrm{Grid}^{\max}V_\mathrm{Grid}(t)\big\},
\end{align}
where $P_\mathrm{IT,0}$ denotes the pre-disturbance IT load power. As shown in Figure~\ref{fig:GridPowerControl}, when $V_\mathrm{Grid}\geq P_\mathrm{IT,0}/I_\mathrm{Grid}^{\max}$, the rectifier maintains its grid-side power draw at $P_\mathrm{IT,0}$ by increasing $I_\mathrm{Grid}$ as the voltage falls. Below this threshold, the rectifier holds $I_\mathrm{Grid}$ at $I_\mathrm{Grid}^{\max}$, and the grid-side power reference $P_\mathrm{Grid}^*$ decreases in proportion to $V_\mathrm{Grid}$.

The power deficit $\Delta P^*$ is calculated as:
\begin{align}
    \Delta P^*(t) &= P_\mathrm{IT,0} - P_\mathrm{Grid}^*\big(V_\mathrm{Grid}(t)\big) \nonumber \\
    &=\max \big\{0, P_\mathrm{IT,0}- I_\mathrm{Grid}^{\max}V_\mathrm{Grid}(t)\big\}.
\end{align}
The power deficit is then allocated to UPS battery discharge power $P_{\mathrm{Bat}}\geq 0$ and IT load curtailment $P_{\mathrm{Cur}}\geq 0$ such that 
\begin{equation}
\label{eq:balance}
    P_\mathrm{Bat}+P_\mathrm{Cur}=\Delta P^*(V_\mathrm{Grid}).
\end{equation}
The actual IT load becomes $P_\mathrm{IT}=P_\mathrm{IT,0}-P_\mathrm{Cur}$.

\subsection{Scenario~1: Battery Support Only}

In this scenario,
the IT load is not curtailed with $P_\mathrm{Cur}=0$, and the battery
supplies the entire power deficit. Upon activation, the rectifier switches
from DC-voltage control to constant-power control and tracks $P_\mathrm{Grid}^*$ in \eqref{eq:Pgridobj}. The
battery DC/DC converter takes over the DC voltage control through a
PI controller, so the battery power adjusts automatically until
$P_\mathrm{Bat}=P_\mathrm{IT}-P_\mathrm{Grid}=\Delta P^*$. The handover
must be completed within $T_\mathrm{hold}$ in \eqref{eq:holdup}. In this way, the IT
load remains unaffected, at the cost of UPS battery discharge energy throughout the sag.

\subsection{Scenario 2: Load Curtailment Only}
This scenario applies to data centers without a UPS battery, or whose
battery is unavailable during the sag. The rectifier keeps its
DC-voltage control, whose current reference is limited to
$I_\mathrm{Grid}^{\max}$; and the IT load is curtailed in software,
e.g., by GPU power capping, to $P_\mathrm{IT}^*$ given by \eqref{eq:PITs2}:
\begin{equation}
\label{eq:PITs2}
    P_\mathrm{IT}^*=\min\{P_\mathrm{IT,0},\,I_\mathrm{Grid}^\mathrm{set}V_\mathrm{Grid}\},
\end{equation}
where $I_\mathrm{Grid}^\mathrm{set}\in [1,I_\mathrm{Grid}^{\max}]$ denotes the grid-side current setpoint in per unit. The lower bound of 1~p.u. ensures that the grid power draw meets the proportional minimum power requirement. In practice, $I_\mathrm{Grid}^\mathrm{set}$ is selected slightly below $I_\mathrm{Grid}^{\max}$ to reserve a rectifier power margin of $(I_\mathrm{Grid}^{\max}-I_\mathrm{Grid}^\mathrm{set})V_\mathrm{Grid}$; this indicates curtailing slightly more IT load than the power deficit ($P_\mathrm{Cur}\!>\!\Delta P^*$), allowing the rectifier to use the reserved margin to recharge the DC-link capacitor and restore its voltage.

IT load curtailment needs to satisfy two conditions. First, the DC-link capacitor supplies the power deficit until load curtailment takes effect. Thus for a step sag, the load curtailment delay $\tau_\mathrm{Cur}$ must  satisfy $\tau_\mathrm{Cur}< T_\mathrm{hold}$, where $T_\mathrm{hold}$ is the time \eqref{eq:holdup} the capacitor can sustain the deficit before $V_\mathrm{DC}$ reaches $V_\mathrm{DC,min}$.
Second, the IT power setpoint in \eqref{eq:PITs2} must stay above the noncurtailable demand
$P_\mathrm{IT,min}$, including latency-critical workloads, storage, and
networking, i.e., $V_\mathrm{Grid}\geq P_\mathrm{IT,min}/I_\mathrm{Grid}^\mathrm{set}$.
Below this voltage, the residual power deficit keeps discharging the DC link, 
and can trip the UPS once $V_\mathrm{DC}$ falls below $V_\mathrm{DC,min}$.

\subsection{Scenario 3: Coordinated Battery and Load Curtailment}

In this scenario, the UPS battery and IT load are coordinated to achieve LVRT, while
the rectifier is operating in a constant power control mode and tracks $P_\mathrm{Grid}^*$ in \eqref{eq:Pgridobj}. The battery regulates $V_\mathrm{DC}$ via a PI controller. The power deficit is optimally
allocated between the battery and load curtailment at minimum control cost by solving the power dispatch problem \eqref{eq:OptimalPro}:
\begin{subequations}
\label{eq:OptimalPro}
\begin{align}
& \min_{\Pb,P_\tc}\ \Cb(\Pb)+C_\tc(P_\tc) \label{eq:costfunction}\\
& \mathrm{s.t.}\quad \Pb+P_\tc=\Delta P^*(V_\mathrm{Grid}),\\
& \qquad\ \Pb\geq 0,\quad 0\leq P_\tc\leq P_\mathrm{IT,0}-P_\mathrm{IT,min}, \label{eq:batcon}
\end{align}
\end{subequations}
where
$C_\mathrm{Bat}(\cdot)$ and $C_\tc(\cdot)$ denote the cost functions of battery
discharge and IT load curtailment. 
Constraint~\eqref{eq:batcon} omits an upper bound on $\Pb$ as the UPS battery is sized to supply the full IT load.
Substituting $\Pb=\Delta P^*-P_\tc$ reduces
\eqref{eq:OptimalPro} to
\begin{align}
    \label{eq:loadcurtail}
    P_\tc^*=\!\!\operatorname*{arg\,min}_{0 \le P_{\mathrm{Cur}} \le \bar{P}_{\mathrm{Cur}}}\!
    \!\!
    \big[\Cb(\Delta P^*\!-\!P_\tc)\!+\!C_\tc(P_\tc)\big],
\end{align}
where $\bar{P}_\tc=\min\{\Delta P^*,\,P_\mathrm{IT,0}-P_\mathrm{IT,min}\}$. 
Since the feasible set is a closed interval and the objective is
continuous, a minimizer exists; if it is not unique, the one with
the least load curtailment is selected. Since $\Delta P^*$ and $\bar{P}_\tc$
depend only on $V_\mathrm{Grid}$, $P_\tc^*$ is a function of $V_\mathrm{Grid}$
and can be computed offline and stored as a lookup table. 

During LVRT operation under a voltage sag, the controller retrieves $P_\tc^*$ from the lookup table for the measured grid voltage, and issues the power setpoint $P_\mathrm{IT}^*=P_\mathrm{IT,0}-P_\tc^*$ to the IT load and $P_\mathrm{Grid}^*$ to the
rectifier. Because the UPS battery
regulates $V_\mathrm{DC}$, it automatically supplies
$P_\mathrm{IT}-P_\mathrm{Grid}$. The two conditions in Scenario~2 thus are not needed. When the grid voltage recovers, the IT power setpoint $P_\mathrm{IT}^*$ returns to its pre-sag value $P_\mathrm{IT,0}$, the rectifier resumes DC-link voltage control, and the battery returns to standby.

\section{Case Studies}
\label{sec:case}
\subsection{Experiment Settings}

The EMT simulation-based case studies use the TAMU-ERCOT PSCAD model of an AI data center~\cite{pscadModel}, which includes the rectifier, DC link, inverter, battery-side DC/DC converter, and IT load.
The settings are listed in Table~\ref{tab:settings}:
\begin{table}[htbp]
\vspace{-10pt}
\caption{Simulation settings.}
\label{tab:settings}
\vspace{-3pt}
\centering\small
\begin{tabular}{p{0.55\columnwidth}p{0.3\columnwidth}}
\toprule
Parameter & Value\\
\midrule
Facility pre-fault demand / IT demand & 1~MW\\
DC-link setpoint / capacitance & 1000~V / 20~mF\\
Minimum DC-link voltage $V_\mathrm{DC, min}$ & 800~V \\
LVRT controller activation threshold & $V_\mathrm{Grid}=0.85$~p.u.\\
Converter current limit & 1.2~p.u.\\
Battery rating & 1~MW \\
$P_\mathrm{IT,min}$ & 0.4~p.u. \\
Curtailment delay &  20 ms \\ 
EMT time step & 25~$\mu$s\\
\bottomrule
\end{tabular}
\end{table}

The conventional UPS tripping mechanism and the three scenarios in
Section~\ref{sec:Scenario} are tested on a single data center facility. The grid voltage follows the
Large Load Model Quality Test (MQT) in ERCOT's Dynamics Working Group
Procedure Manual~\cite{ercotMQT}, which ERCOT uses to verify the
ride-through capability of large computational loads. In this test, the facility is
connected to an infinite bus whose voltage follows a profile consisting
of a slow voltage ramp, dips, and shallow and deep faults
covering the bands in Table~\ref{tab:vrt}. The facility must remain
connected throughout and satisfy the minimum power requirement, shown as $P_\mathrm{Req}$ in the results.
\subsection{Tripping Case without LVRT Control}

\begin{figure}[htbp]
    \centering
\includegraphics[width=0.9\linewidth]{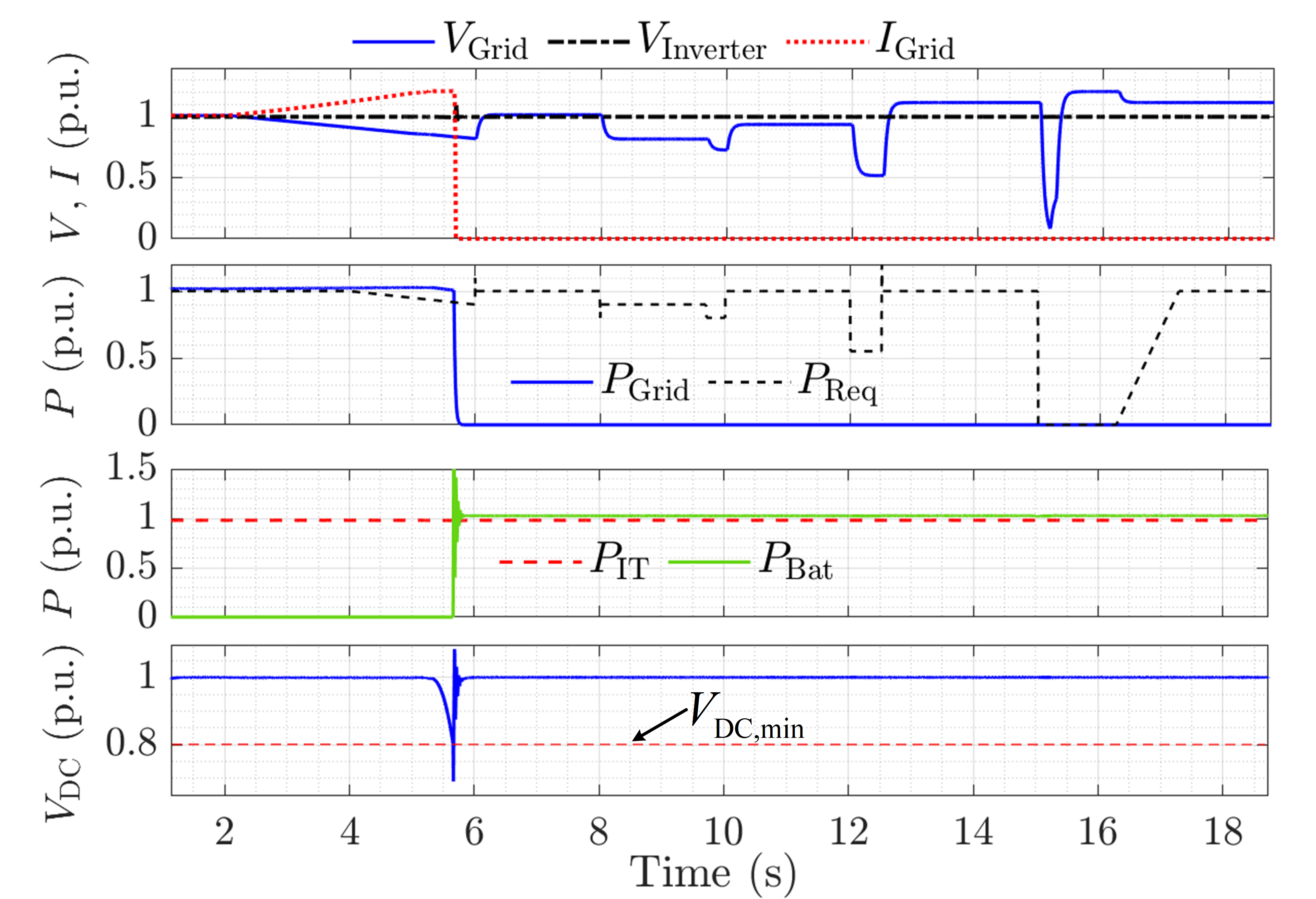}
\vspace{-5pt}
    \caption{Tripping case without LVRT control.}
    \label{fig:case1}
    \vspace{-5pt}
\end{figure}

First, the conventional UPS operation and protection mechanism is tested under the MQT curve, i.e., once $V_\mathrm{DC} < V_\mathrm{DC, min}$,  the data center is disconnected from the grid and the battery is used to fully support the IT load. The result is shown in Figure~\ref{fig:case1}. At $t=5.3$~s, $V_\mathrm{Grid}$ drops to 0.83~p.u. and keeps dropping. The rectifier raises $I_\mathrm{Grid}$ to its limit of 1.2~p.u., but $I_\mathrm{Grid}^{\max}V_\mathrm{Grid}$ is below $P_\mathrm{IT}$, hence, $V_\mathrm{DC}$ collapses. Once $V_\mathrm{DC}<V_\mathrm{DC,min}$, the rectifier is tripped, and the battery takes over the entire load. The IT load source voltage $V_\mathrm{Inverter}$ has little interruption; hence, $P_\mathrm{IT}$ remains constant. However,  $P_\mathrm{Grid}$ drops from 1~p.u. to zero during a sag and does not satisfy the ride-through requirement.

\subsection{Scenario 1: Battery Support Only}

\begin{figure}[htbp]
    \centering
\includegraphics[width=0.9\linewidth]{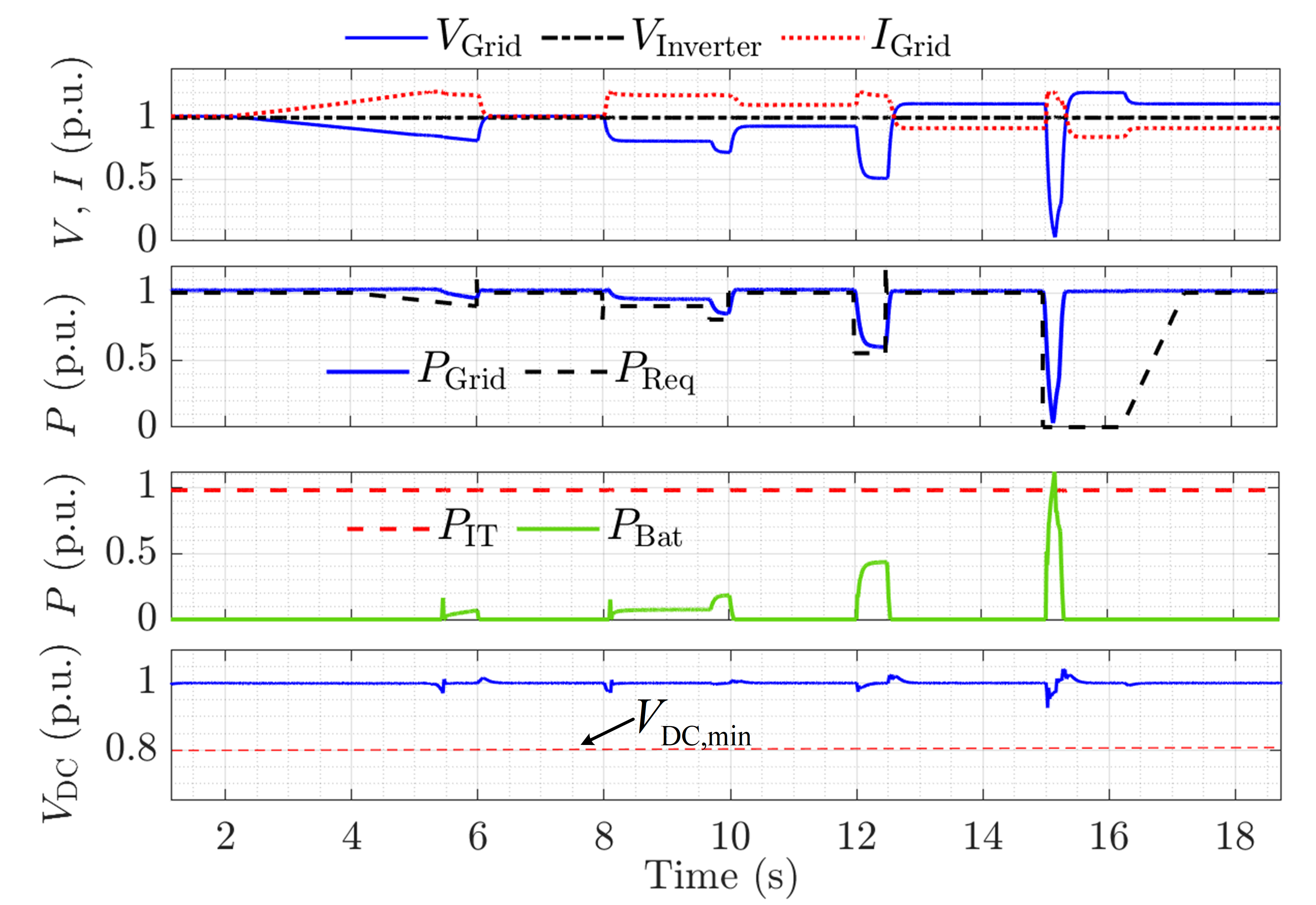}
\vspace{-5pt}
    \caption{LVRT test result. (Scenario 1: Battery support only)}
    \label{fig:Case2}
    \vspace{-5pt}
\end{figure}

Figure~\ref{fig:Case2} shows battery-only support under the MQT profile. While
$V_\mathrm{Grid}\geq P_\mathrm{IT,0}/I_\mathrm{Grid}^{\max}$, the rectifier keeps constant
power and $I_\mathrm{Grid}$ rises as the voltage falls. Once the current
reaches $I_\mathrm{Grid}^{\max}=1.2$~p.u., $P_\mathrm{Grid}$ follows $I_\mathrm{Grid}^{\max}V_\mathrm{Grid}$,
e.g., about 0.84~p.u. at 0.7~p.u. and 0.6~p.u. at 0.5~p.u., and the battery
automatically supplies the deficit. The IT
load is never curtailed, and $V_\mathrm{DC}$ stays within $\pm$5\% of its
rating, with only brief transients at sag onset and recovery, so no trip
occurs. Throughout, $P_\mathrm{Grid}$ remains above
$P_\mathrm{Req}$, and returns to its pre-fault value as soon as the
voltage recovers.

\subsection{Scenario 2: Load Curtailment Only}
\begin{figure}[htbp]
    \centering
       \vspace{-10pt}
\includegraphics[width=0.9\linewidth]{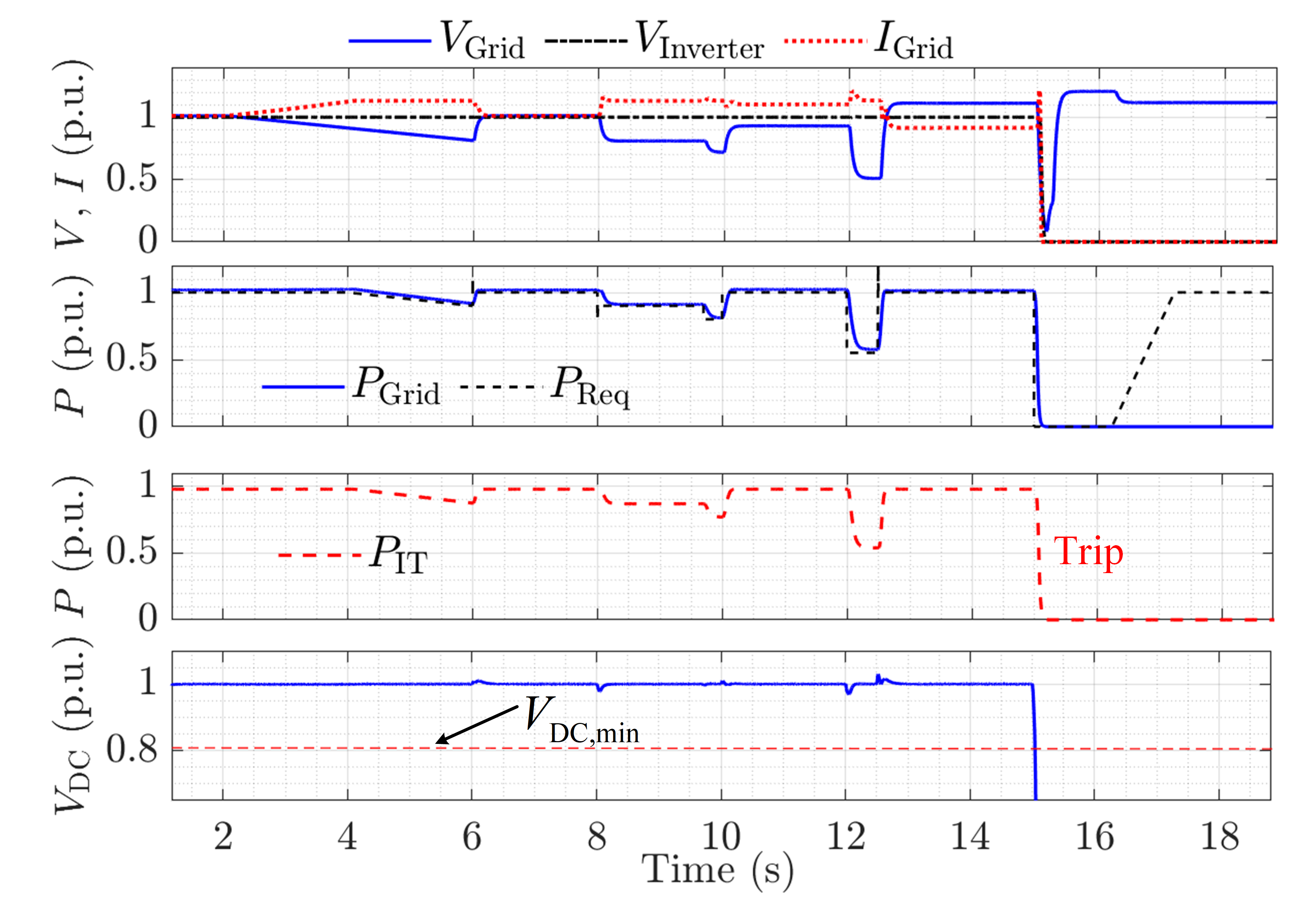}
\vspace{-5pt}
    \caption{LVRT test result. (Scenario 2: Load curtailment only)}
    \label{fig:Case3}
    \vspace{-5pt}
\end{figure}
Without the battery, curtailment alone rides through all sags above
$P_\mathrm{IT,min}/I_\mathrm{Grid}^\mathrm{set}$ but not the deepest one, as shown in Figure~\ref{fig:Case3}.
To leave the rectifier a margin for regulating
$V_\mathrm{DC}$ against the curtailment delay and losses, $I_\mathrm{Grid}^\mathrm{set}$
is set to 1.1~p.u., below $I_\mathrm{Grid}^{\max}$. For all sags with
$V_\mathrm{Grid}>P_\mathrm{IT,min}/I_\mathrm{Grid}^\mathrm{set}\approx0.36$~p.u.,
$P_\mathrm{IT}$ is reduced to $I_\mathrm{Grid}^\mathrm{set}V_\mathrm{Grid}$, the
rectifier keeps $V_\mathrm{DC}$ near its rating, and $P_\mathrm{Grid}$
stays above $P_\mathrm{Req}$. During the zero-voltage fault, the
grid supplies no power, while the IT load cannot be reduced below
$P_\mathrm{IT,min}$. The residual deficit discharges the DC link, and
the facility trips once $V_\mathrm{DC}$ reaches $V_\mathrm{DC,min}$.
\subsection{Scenario 3: Coordinated Control}
\begin{figure}[tbp]
    \centering
 \includegraphics[width=0.9\linewidth]{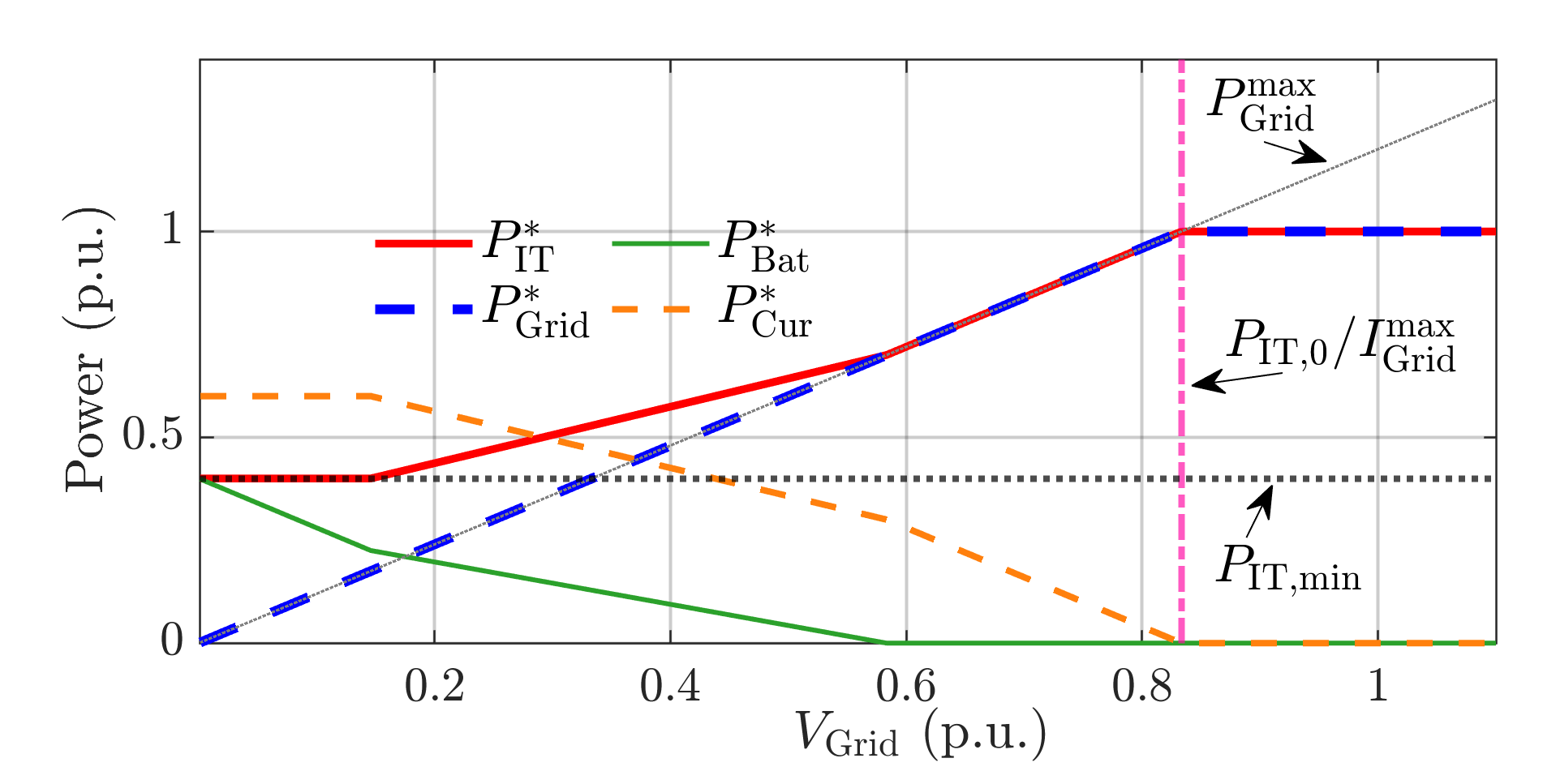}
      \vspace{-5pt}
    \caption{Control reference versus grid voltage in Scenario~3.}
     \vspace{-5pt}
    \label{fig:PPrefVgrid}
\end{figure}
\begin{figure}[tbp]
    \centering
\includegraphics[width=0.9\linewidth]{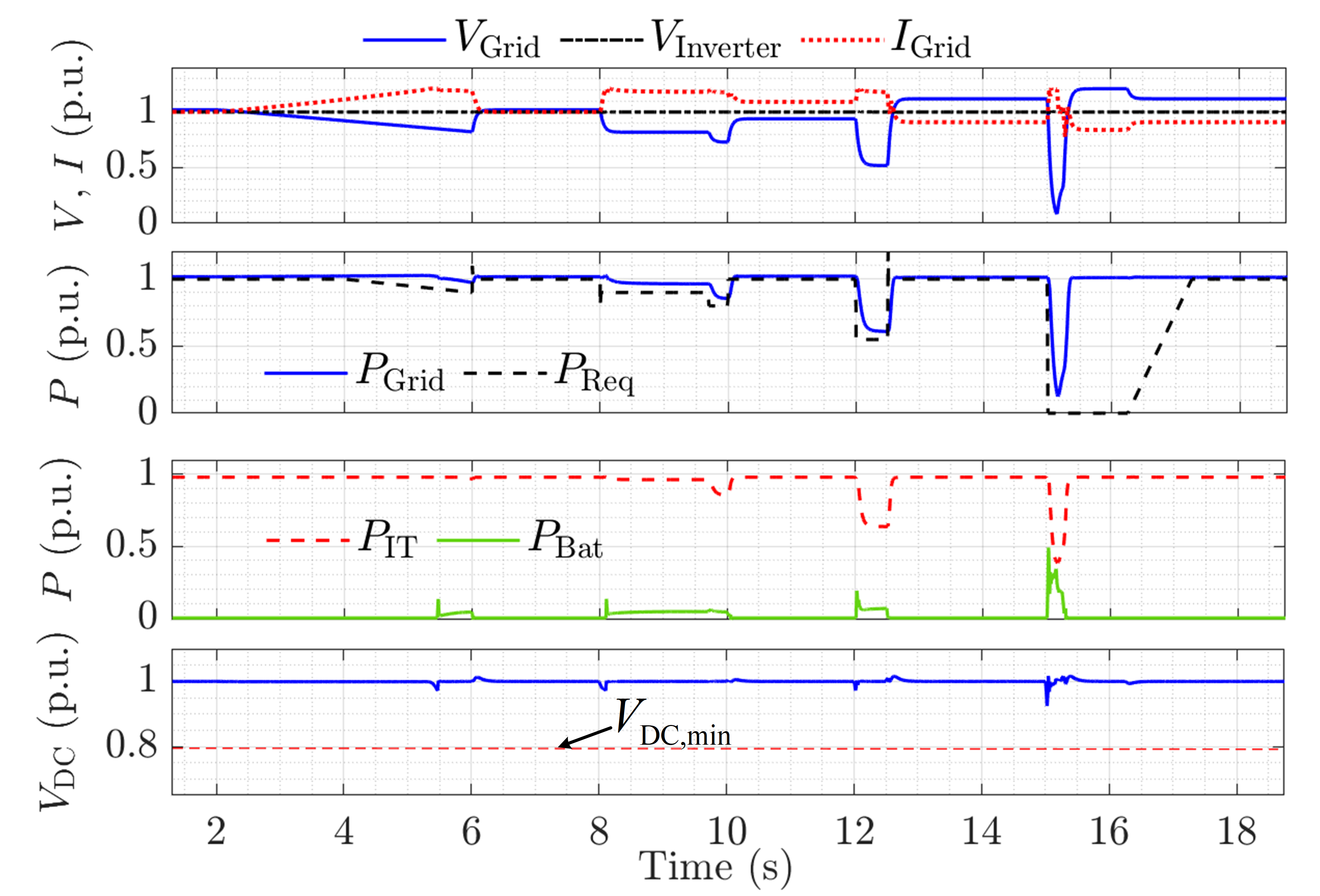}
   \vspace{-5pt}
    \caption{LVRT test result. (Scenario~3: Coordinated control)}
    \vspace{-10pt}
    \label{fig:Case4}
\end{figure}

Quadratic cost functions in \eqref{eq:costfunction} are considered:
\begin{align}
     \Cb = \frac{1}{2} \alpha_\mathrm{Bat} \Pb^2 + \beta_\mathrm{Bat}  P_\mathrm{Bat}, \nonumber\\
    C_\tc = \frac{1}{2} \alpha_\tc P_\tc^2 + \beta_\tc  P_\tc.
\end{align}
Accordingly,
the optimal load curtailment is given by:
\begin{equation}
    P_\tc^*=\left[ \frac{\alpha_\mathrm{Bat} \Delta P^* + \beta_\mathrm{Bat} - \beta_\tc }{ \alpha_\mathrm{Bat} +  \alpha_\tc} \right]_0^{\bar{P}_\tc}
\end{equation}
where $\left[ x \right]_l^u = \min\{u, \max\{l,x\}\}$.
With $ \alpha_\mathrm{Bat}=0.4,\beta_\mathrm{Bat}=0.1,  \alpha_\tc=0.3, \beta_\tc=0.01$, and $P_\mathrm{IT,min}=0.4$~p.u., references $P_\mathrm{Grid}^*, P_\mathrm{IT}^*, \Pb^*$ and $P_\tc^*$ are shown in Figure~\ref{fig:PPrefVgrid}. It can be observed that 
load curtailment has a lower marginal cost at small depths ($\beta_\tc<\beta_\mathrm{Bat}$) but a limited range. Three regimes appear as the sag deepens. For $0.583\leq V_\mathrm{Grid}<0.833$~p.u., the deficit is covered by curtailment alone.
For $0.146\leq V_\mathrm{Grid}<0.583$~p.u., 
the battery and curtailment share each further increment of the deficit.
With the IT load at $P_\mathrm{IT,min}$, curtailment remains at its
limit, and the battery supplies any further increase in the deficit.

Figure~\ref{fig:Case4} shows the cost-optimal coordination test result with the reference power set according to Figure~\ref{fig:PPrefVgrid}. It can be observed that in the
0.5-p.u. sag, $P_\mathrm{IT}$ is reduced only to about 0.64~p.u., and the battery supplies the remaining power deficit.

\section{Conclusion}
This paper formulates data center LVRT as a DC-link power balance problem and proposes control strategies that coordinate UPS battery and computing load. Under current-limited operation, the rectifier draws grid power in proportion to the grid voltage. The proposed control allocates the remaining power deficit between battery discharge and load curtailment to minimize total control cost.
EMT simulations show that
the conventional UPS strategy disconnects the facility even under a mild voltage
sag. With battery support or the coordinated control, the facility stays
connected through all tested sags, including a near-zero-voltage fault,
with a stable DC link, uninterrupted IT load, and grid power draw above
the ERCOT requirement. Load curtailment alone requires no battery but cannot ride through the deepest voltage sag.
\bibliographystyle{IEEEtran}
\bibliography{IEEEabrv,mybibfileGM2027WXY}
\end{document}